\documentclass[11pt,letterpaper]{myarticle}
\let\section=\subsection     
\let\subsection=\subsubsection
\renewcommand\thesubsection{\arabic{subsection}}     
\usepackage{anysize,graphicx,amsfonts,mathrsfs,latexsym,amsmath,
amsopn} 
\usepackage{feynmp}
\newcommand{\Lag}{\ensuremath{\mathscr{L}}}
\newcommand{\dyad}[1]{\ensuremath{\overset{\leftrightarrow}{#1}}}
\newcommand{\funcd}[2]{\frac{\delta #1}{\delta #2}}
\renewcommand{\i}{\ensuremath{\mathrm{i}}}

\marginsize{2.5cm}{2.5cm}{2.5cm}{2.5cm}

\begin{document}
\begin{center}
   {\large \bf Finite pion width effects on the rho-meson }\\[1mm]
   {\large \bf and di-lepton spectra}\\[3mm]
   Hendrik van Hees\footnote{e-mail:h.vanhees@gsi.de} 
        and J\"o{}rn Knoll \footnote{e-mail:j.knoll@gsi.de}\\[3mm]
   {\small \it  Gesellschaft f\"ur Schwerionenforschung (GSI) \\
   Planckstr. 1, D-64291 Darmstadt, Germany \\[3mm] }
\end{center}

\begin{abstract}\noindent
  Within a field theoretical model where all damping width effects are
  treated self-consistently we study the changes of the spectral
  properties of $\rho$-mesons due to the finite damping width of the
  pions in dense hadronic matter at finite temperature. The
  corresponding effects in the di-lepton yields are presented. Some
  problems concerning the self consistent treatment of vector or gauge
  bosons are discussed.
\end{abstract}

\section{Introduction}
The properties of vector mesons in a dense hadronic medium have
attracted much attention in recent times.  Measurements of di-leptons
in nuclear collisions promise to access such properties
experimentally. Recent experiments by the CERES and DLS collaborations
\cite{ceres95,ceres95-2,dls97} show interesting features in the low
lepton pair mass spectrum between $300$ to $600\,\text{MeV}$. Various
effects which change the mass and/or the width, or brief the spectral
properties of the vector mesons in dense matter have been explored to
explain the observed enhancement seen in heavy projectile-target
collisions compared to proton-proton data.  High resolution
experiments with the upgrade of CERES and the new di-lepton project
HADES at GSI will sharpen the view on the spectral information of
vector mesons.

In most of the theoretical investigations the damping width attained
by the asymptotically stable particles in the dense matter environment
has been ignored sofar. In this contribution we study the in-medium
properties of the $\rho$-meson due to the damping width of the pions
in a dense hadron gas within a self consistent scheme.
 
\section{The model}

In order to isolate the pion width effects we discard baryons and
consider a purely mesonic model system consisting of pions,
$\rho$-mesons, and for curiosity also the chiral partner of the
$\rho$-, the $a_1$-meson with the interaction Lagrangian
\begin{equation}\label{Lint}
\Lag^{\text{int}}=g_{\rho\pi\pi}\rho_{\mu} \pi^* \dyad{\partial}{}^{\mu} 
\pi+g_{\pi\rho a_1}\pi\rho_{\mu} a_1^{\mu}+\frac{g_{\pi 4}}{8}
(\pi^* \pi)^2.
\end{equation}
The first two coupling constants are adjusted to provide the
corresponding vacuum widths of the $\rho$- and $a_1$-meson at the
nominal masses of $770$ and $1200\,\text{MeV}$ of
$\Gamma_{\rho}=150\,\text{MeV}$ and $\Gamma_{a_1}=400\,\text{MeV}$,
respectively. The four-pion interaction is used as a tool to furnish
additional collisions among the pions. The idea of this term is to
provide pion damping widths of $50\,\text{MeV}$ or more as they would
occur due to the strong coupling to the $NN^{-1}$ and $\Delta N^{-1}$
channels in an environment at finite baryon density.

The $\Phi$-functional method originally proposed by Baym\cite{baym62}
provides a self-consistent scheme applicable even in the case of broad
resonances. It bases on a re-summation for the partition sum
\cite{lw60,cjt74}.  Its two particle irreducible part $\Phi[G]$
generates the irreducible self-energy $\Sigma(x,y)$ via a functional
variation with respect to the propagator $G(y,x)$, i.e.
\begin{equation}
\label{varphi}
-\i \Sigma (x,y) =\funcd{\i\Phi}{\i G(y,x)}.  
\end{equation}
Thereby $\Phi$ solely depends on fully re-summed, i.e.
self-consistently generated propagators $G(x,y)$. In graphical terms,
the variation (\ref{varphi}) with respect to $G$ is realized by
opening a propagator line in all diagrams of $\Phi$. Further details
and the extension to include classical fields or condensates into the
scheme are given in ref.~\cite{kv97}.

Truncating $\Phi$ to a limited subset of diagrams, while preserving
the variational relation (\ref{varphi}) between
$\Phi^{\text{(appr.)}}$ and $\Sigma^{\text{(appr.)}}(x,y)$ defines an
approximation with built-in consistency.  Baym\cite{baym62} showed
that such a scheme is conserving at the expectation value level of
conserved currents related to global symmetries of the original
theory, that its physical processes fulfill detailed balance and
unitarity and that at the same time the concept is thermodynamically
consistent. However symmetries and conservation laws may no longer be
maintained on the correlator level, a draw-back that will lead to
problems for the self-consistent treatment of vector and gauge
particles on the propagator level, as discussed in sect. 3.

Interested in width effects, we drop changes in the real parts of the
self energies. This entitles to drop tadpole contributions for the
self energies. For our model Lagrangian (\ref{Lint}) one obtains the
following diagrams for $\Phi$ at the two-point level which generate
the subsequently given three self energies $\Pi_{\rho}$, $\Pi_{a_1}$
and $\Sigma_{\pi}$

\def\fmfsdot#1{\fmfv{decor.shape=circle,decor.filled=full,decor.size=1.2thick}
{#1}}
\def\GPhipirhopi{
\parbox{12mm}{
\begin{fmfgraph*}(12,0)
\fmfpen{thick}
\fmfleft{l}
\fmfright{r}
\fmfforce{(0.0w,0.5h)}{l}
\fmfforce{(1.0w,0.5h)}{r}
\fmf{gluon,left=.1,label=$\rho$,l.d=7.5,label.side=right}{r,l}
\fmf{plain,left=.9,tension=.5,label=$\pi$}{l,r}
\fmf{plain,left=.9,tension=.5,label=$\pi$,l.side=right}{r,l}
\fmfsdot{l,r}
\end{fmfgraph*}
}}
\def\GPhipirhoa{
\parbox{12mm}{
\begin{fmfgraph*}(12,0)
\fmfpen{thick}
\fmfleft{l}
\fmfright{r}
\fmfforce{(0.0w,0.5h)}{l}
\fmfforce{(1.0w,0.5h)}{r}
\fmf{plain,left=.9,tension=.5,label=$\pi$}{l,r}
\fmf{gluon,left=.1,label=$\rho$,l.d=7.5,l.side=right}{r,l}
\fmf{photon,left=.9,tension=.3,l.d=3.,label=$a_1$,l.side=right}{r,l}
\fmfsdot{l,r}
\end{fmfgraph*}
}}
\def\GPhipi{
\parbox{12mm}{
\begin{fmfgraph*}(12,0)
\fmfpen{thick}
\fmfleft{l}
\fmfright{r}
\fmfforce{(0.0w,0.5h)}{l}
\fmfforce{(1.0w,0.5h)}{r}
\fmf{plain,left=.9,tension=.5,l.d=1.5,label=$\pi$,l.side=right}{r,l}
\fmf{plain,left=.3,tension=.5,l.d=1.8,label=$\pi$,l.side=right}{r,l}
\fmf{plain,left=.3,tension=.5,l.d=1.7,label=$\pi$}{l,r}
\fmf{plain,left=.9,tension=.5,l.d=1.3,label=$\pi$}{l,r}
\fmfsdot{l,r}
\end{fmfgraph*}}
}
\def\GPipipi{
\parbox{20mm}{
\begin{fmfgraph*}(20,10)
\fmfpen{thick}
\fmfleft{l}
\fmfright{r}
\fmfforce{(0.0w,0.5h)}{l}
\fmfforce{(1.0w,0.5h)}{r}
\fmfforce{(0.2w,0.5h)}{ol}
\fmfforce{(0.8w,0.5h)}{or}
\fmf{plain,left=.7,label=$\pi$,l.s=right}{or,ol}
\fmf{plain,left=.7,label=$\pi$,l.s=right}{ol,or}
\fmf{gluon}{l,ol}
\fmf{gluon}{or,r}
\fmfsdot{ol,or}
\end{fmfgraph*}
}}
\def\GPipia{
\parbox{20mm}{
\begin{fmfgraph*}(20,10)
\fmfpen{thick}
\fmfleft{l}
\fmfright{r}
\fmfforce{(0.0w,0.5h)}{l}
\fmfforce{(1.0w,0.5h)}{r}
\fmfforce{(0.2w,0.5h)}{ol}
\fmfforce{(0.8w,0.5h)}{or}
\fmf{photon,left=.9,tension=.5,l.d=3.5,label=$a_1$,l.s=right}{or,ol}
\fmf{plain,left=.7,tension=.5,label=$\pi$,l.s=right}{ol,or}
\fmf{gluon}{l,ol}
\fmf{gluon}{or,r}
\fmfsdot{ol,or}
\end{fmfgraph*}
}}
\def\GPipirho{
\parbox{20mm}{
\begin{fmfgraph*}(20,10)
\fmfpen{thick}
\fmfleft{l}
\fmfright{r}
\fmfforce{(0.0w,0.5h)}{l}
\fmfforce{(1.0w,0.5h)}{r}
\fmfforce{(0.2w,0.5h)}{ol}
\fmfforce{(0.8w,0.5h)}{or}
\fmf{gluon,right=.5,label=$\rho$,l.s=left,l.dist=3}{ol,or}
\fmf{plain,left=.8,label=$\pi$,l.s=right,l.dist=3}{ol,or}
\fmf{photon}{l,ol}
\fmf{photon}{or,r}
\fmfsdot{ol,or}
\end{fmfgraph*}
}}
\def\GSigpirho{
\parbox{20mm}{
\begin{fmfgraph*}(20,10)
\fmfpen{thick}
\fmfleft{l}
\fmfright{r}
\fmfforce{(0.0w,0.5h)}{l}
\fmfforce{(1.0w,0.5h)}{r}
\fmfforce{(0.2w,0.5h)}{ol}
\fmfforce{(0.8w,0.5h)}{or}
\fmf{gluon,right=.5,label=$\rho$,l.side=left,l.dist=3}{ol,or}
\fmf{plain,right=.8,label=$\pi$,l.side=left,l.dist=3}{or,ol}
\fmf{plain}{l,ol}
\fmf{plain}{or,r}
\fmfsdot{ol,or}
\end{fmfgraph*}
}}
\def\GSigarho{
\parbox{20mm}{
\begin{fmfgraph*}(20,10)
\fmfpen{thick}
\fmfleft{l}
\fmfright{r}
\fmfforce{(0.0w,0.5h)}{l}
\fmfforce{(1.0w,0.5h)}{r}
\fmfforce{(0.2w,0.5h)}{ol}
\fmfforce{(0.8w,0.5h)}{or}
\fmf{photon,left=0.9,label=$a_1$,l.side=right,l.dist=3}{ol,or}
\fmf{gluon,right=.5,label=$\rho$,l.side=left}{ol,or}
\fmf{plain}{l,ol}
\fmf{plain}{or,r}
\fmfsdot{ol,or}
\end{fmfgraph*}
}}
\def\GSigpi{
\parbox{20mm}{
\begin{fmfgraph*}(20,10)
\fmfpen{thick}
\fmfleft{l}
\fmfright{r}
\fmfforce{(0.0w,0.5h)}{l}
\fmfforce{(1.0w,0.5h)}{r}
\fmfforce{(0.2w,0.5h)}{ol}
\fmfforce{(0.8w,0.5h)}{or}
\fmf{plain,left=.7,label=$\pi$,l.s=right}{or,ol}
\fmf{plain,left=.7,label=$\pi$,l.s=left}{ol,or}
\fmf{plain,label=$\pi$}{or,ol}
\fmf{plain}{l,ol}
\fmf{plain}{or,r}
\fmfsdot{ol,or}
\end{fmfgraph*}
}}
\def\Gcross#1#2{\fmfiv{d.sh=cross,d.ang=#1,d.size=10thick}{#2}}
\def\Gjmu{\parbox{15mm}
{\begin{fmfgraph*}(14,15)
\fmfpen{thick}
\fmfright{r}
\fmfleft{i1}
\fmf{gluon,left=.1,label=$\rho$}{i1,v1}
\fmf{phantom,tension=0.4}{v1,r}
\fmffreeze
\fmf{plain,left,label=$\pi$}{v1,r}
\fmf{plain,left}{r,v1}
\fmfsdot{v1}
\end{fmfgraph*}
}}
\def\GPhijmu{\parbox{20mm}
{\begin{fmfgraph*}(20,15)
\fmfpen{thick}
\fmfright{r}
\fmfleft{i1}
\fmf{gluon,left=.1,label=$\rho$}{i1,v1}
\fmf{phantom}{v1,r}
\fmffreeze
\fmf{plain,left,label=$\pi$}{v1,r}
\fmf{plain,left}{r,v1}
\fmfsdot{v1}
\fmffreeze
\Gcross{0}{vloc(__i1)}
\end{fmfgraph*}
}}
\def\GSigjmu{\parbox{10mm}
{\begin{fmfgraph*}(10,15)
\fmfpen{thick}
\fmfright{r}
\fmfleft{l}
\fmftop{t}
\fmfbottom{b}
\fmf{plain}{l,v,r}
\fmf{gluon,left=.1,label=$\rho$}{v,t}
\fmf{phantom}{v,b}
\fmfsdot{v}
\Gcross{0}{vloc(__t)}
\end{fmfgraph*}
}}
\def\PiVert{\parbox{20mm}
{\begin{fmfgraph*}(20,7)
\fmfpen{thick}
\fmfstraight
\fmfright{r}
\fmfleft{l}
\fmftop{t1,t2,t3}
\fmfbottom{b1,b2,b3}
\fmf{gluon,left=.1,tension=1,label=$\rho$}{v,r}
\fmfpoly{shaded,tension=0.7}{v,t2,b2}
\fmf{plain}{t2,vl,b2}
\fmf{gluon,left=.1,tension=3,label=$\rho$}{l,vl}
\fmfsdot{v,vl,t2,b2}
\end{fmfgraph*}
}}
\def\GVert{\parbox{17mm}
{\begin{fmfgraph*}(17,10)
\fmfpen{thick}
\fmfstraight
\fmfright{r1,r2,r3,r,r4,r5,r6}
\fmfleft{l}
\fmftop{t1,t2,t3,t4}
\fmfbottom{b1,b2,b3,b4}
\fmf{phantom}{t1,vt}
\fmf{phantom,tension=0.7}{vt,r4}
\fmf{phantom}{b1,vb}
\fmf{phantom,tension=0.7}{vb,r3}
\fmffreeze
\fmf{gluon,left=.1,tension=0.2,label=$\rho$}{v,r}
\fmfpoly{shaded,tension=0.7}{v,vt,vb}
\fmf{plain}{t1,vt}
\fmf{plain}{b1,vb}
\fmfsdot{v,t1,vt,b1,vb}
\end{fmfgraph*}
}}
\def\GKVert{\parbox{8mm}
{\begin{fmfgraph*}(8,8)
\fmfpen{thick}
\fmftop{t0,t1,t2}
\fmfbottom{b0,b1,b2}
\fmf{plain}{t0,t1,t2}
\fmf{plain}{b0,b1,b2}
\fmf{plain,left=.4}{t1,b1,t1}
\fmfsdot{t1,b1}
\end{fmfgraph*}
}}
\def\GVertLp{\parbox{24mm}
{\begin{fmfgraph*}(24,10)
\fmfpen{thick}
\fmfstraight
\fmfright{r1,r2,r3,r,r4,r5,r6}
\fmfleft{l}
\fmftop{t0,t1,t2,t3,t4}
\fmfbottom{b0,b1,b2,b3,b4}
\fmf{phantom}{t0,v1,vt}
\fmf{phantom}{t1,v1}
\fmf{phantom}{b1,v2}
\fmf{phantom,tension=0.7}{vt,r4}
\fmf{phantom}{b0,v2,vb}
\fmf{phantom,tension=0.7}{vb,r3}
\fmffreeze
\fmf{gluon,left=.1,tension=0.2,label=$\rho$}{v,r}
\fmfpoly{shaded,tension=0.7}{v,vt,vb}
\fmf{plain}{t0,v1,vt}
\fmf{plain}{b0,v2,vb}
\fmf{plain,left=.4}{v1,v2,v1}
\fmfsdot{t0,b0,v,v1,vt,v2,vb}
\end{fmfgraph*}
}}
\def\GVertV{\parbox{10mm}
{\begin{fmfgraph*}(10,10)
\fmfpen{thick}
\fmfstraight
\fmfleft{l1,l2}
\fmfright{r}
\fmf{plain,tension=0.35}{l1,v,l2}
\fmf{gluon,left=.1,label=$\rho$}{v,r}
\fmfsdot{l1,v,l2}
\end{fmfgraph*}
}}

\unitlength=1mm
\begin{fmffile}{kbd}
\fmfset{thick}{1.5pt}
\begin{equation}\label{Phi_Dyson}
\begin{array}{rccccc}
\Phi=&\GPhipirhopi &+&\GPhipirhoa &+&\GPhipi \\[0.8cm]
\Pi_{\rho}=&\GPipipi &+&\GPipia \\[0.8cm]
\Pi_{a_1}=&&&\GPipirho \\[0.8cm]
\Sigma_{\pi}=&\GSigpirho &+&\GSigarho &+&\GSigpi 
\end{array}
\end{equation}
\begin{figure}
\begin{picture}(140,60)
\put(-6,60){
\includegraphics[width=6cm,height=7.4cm,angle=-90]{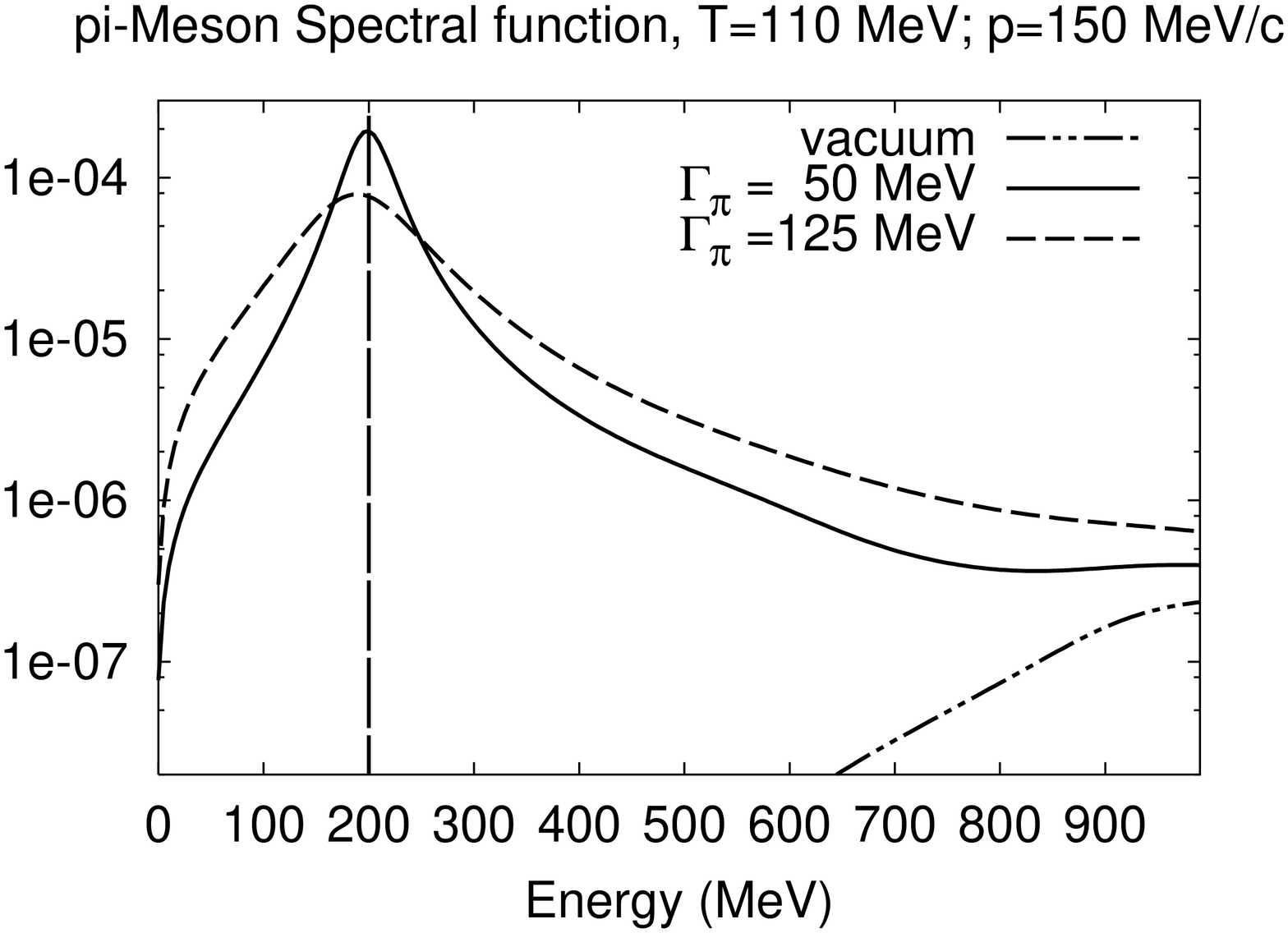}}
\put(68,60){
\includegraphics[width=6cm,height=7.4cm,angle=-90]{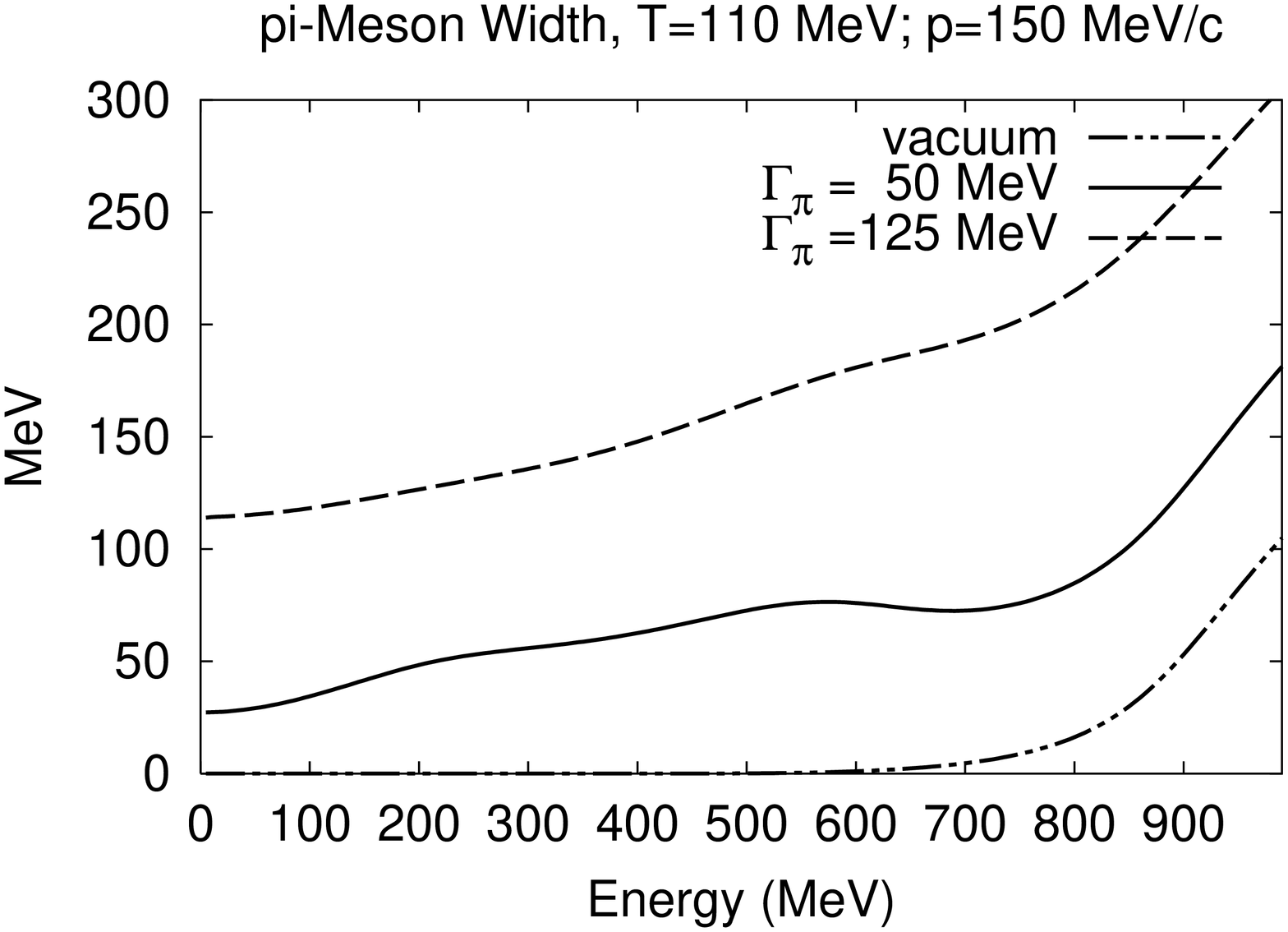}}
\end{picture}
\caption{\textit{Spectral function (left) and decay width (right) of the pion
    as a function of the pion energy at a pion momentum of $150\;
    \text{MeV}/c$ in the vacuum and for two self consistent cases
    discussed in the text.}}
\end{figure}
They are the driving terms for the corresponding three Dyson
equations, which have to be solved self consistently. The above
coupled scheme pictorially illustrates the concept of Newton's
principle of \emph{actio = reactio} and detailed balance provided by
the $\Phi$-functional.  If the self energy of one particle is modified
due to the coupling to other species, these other species also obtain
a complementary term in their self energy.  In vacuum the $\rho$- and
$a_1$-meson have the standard thresholds at $\sqrt{s}=2m_{\pi}$ and at
$3 m_{\pi}$ respectively. For the pion as the only stable particle in
vacuum with a pole at $m_{\pi}$ a decay channel opens at $\sqrt{s}=3
m_{\pi}$ due to the first and last diagram of $\Sigma_{\pi}$.
Correspondingly the vacuum spectral function of the pion shows already
some spectral strength for $\sqrt{s}>3 m_{\pi}$, c.f. fig.~1 (left).

\unitlength=1mm
\begin{figure}[h]
\begin{picture}(118,120)
\put(0,65){
\put(-8,74){{
\includegraphics[width=8cm,height=8cm,angle=-90]{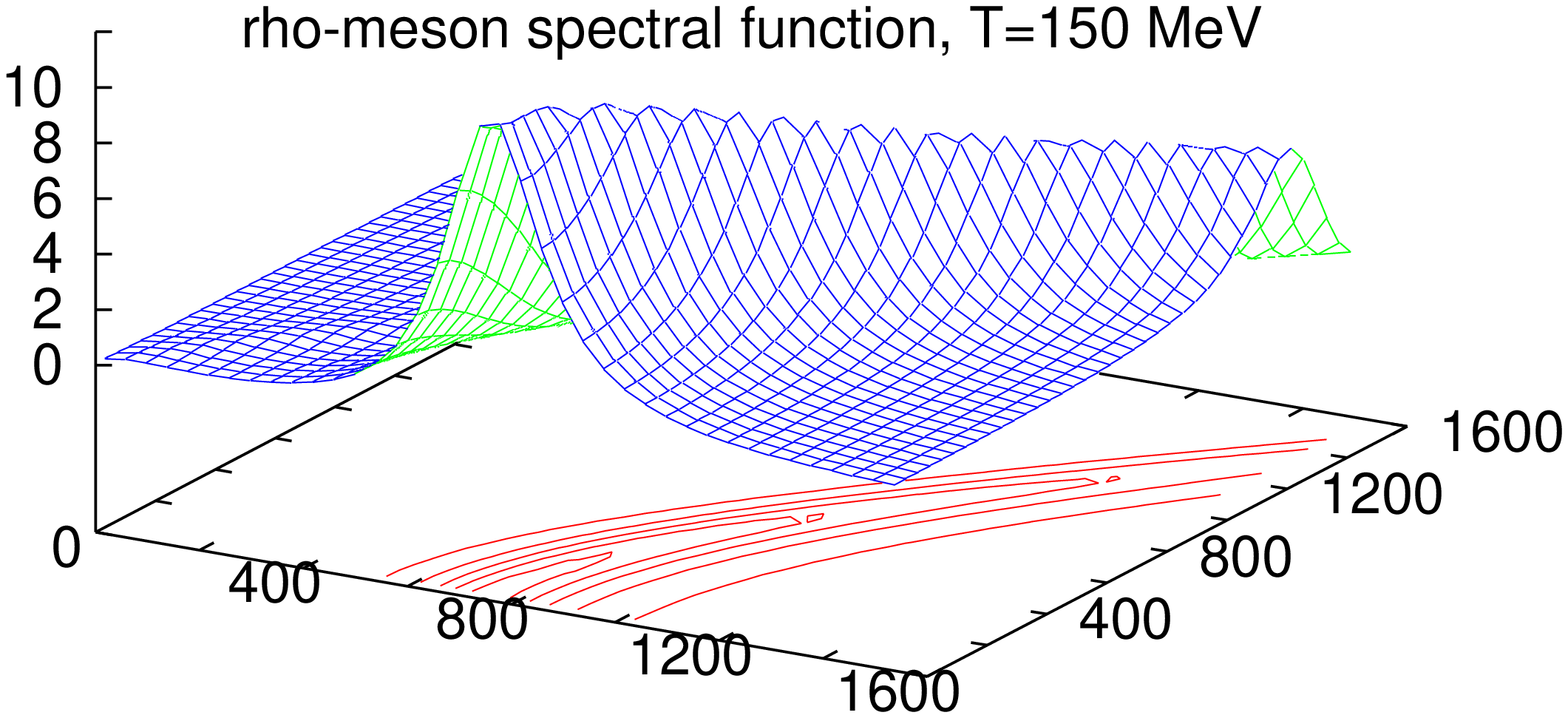}}}
\put(18,5){$p_0$}
\put(60,12){$|\vec p\,|$}
\put(63,55){{
\includegraphics[width=6cm,height=8cm,angle=-90]{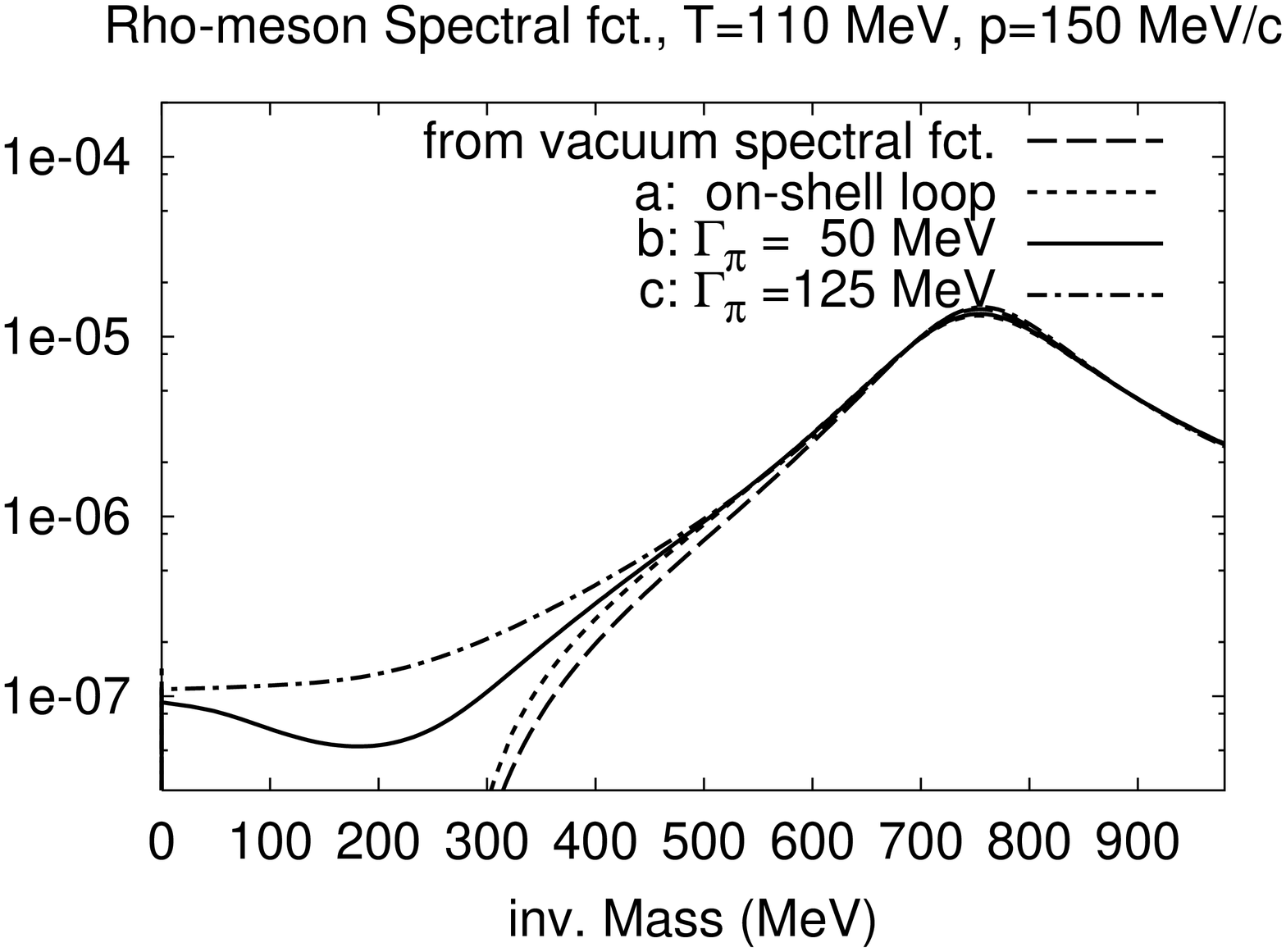}}}
}
\put(-6,60){{
\includegraphics[width=6cm,height=7.8cm,angle=-90]{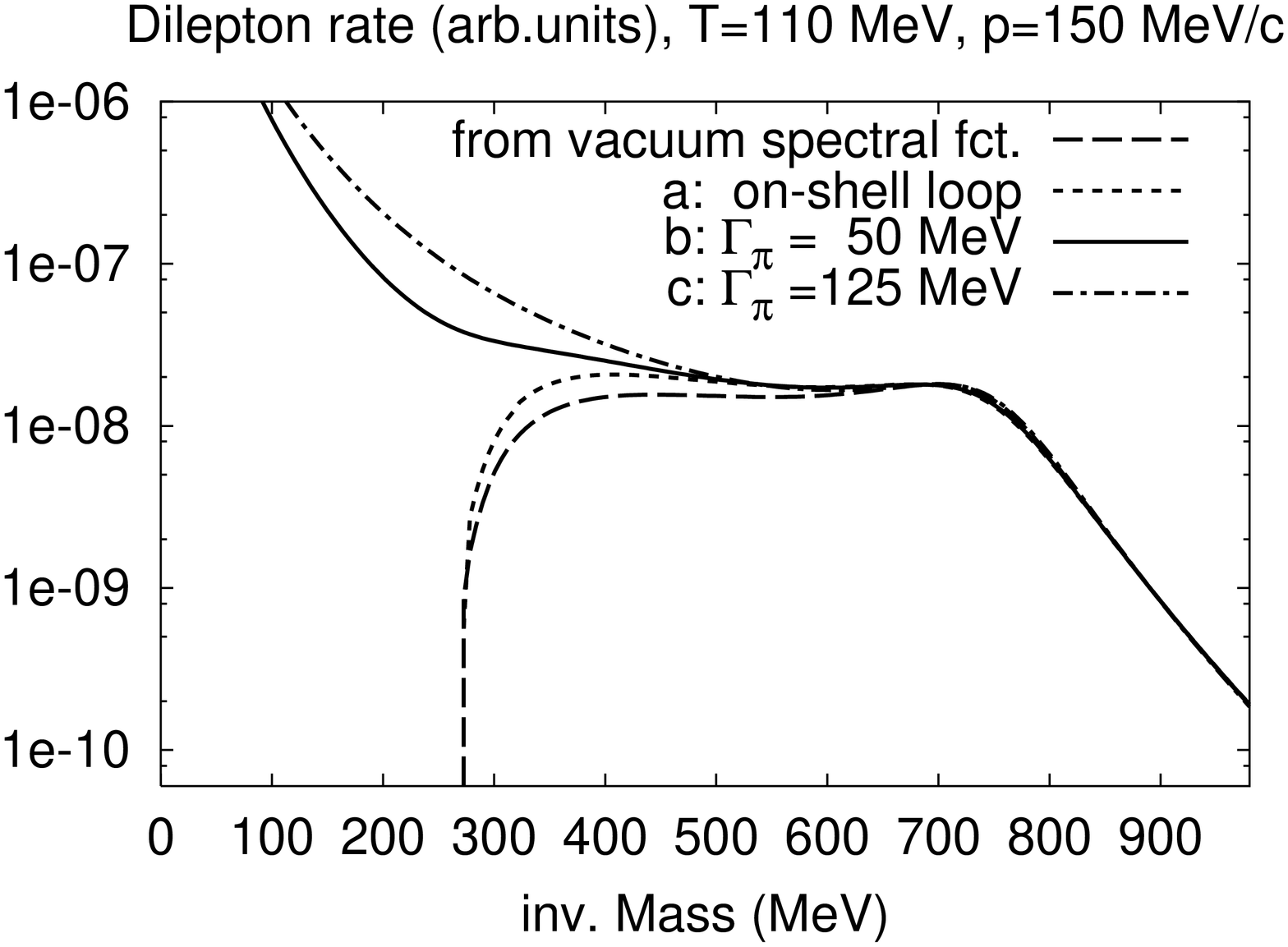}}}
\put(65,60){{
\includegraphics[width=6cm,height=7.8cm,angle=-90]{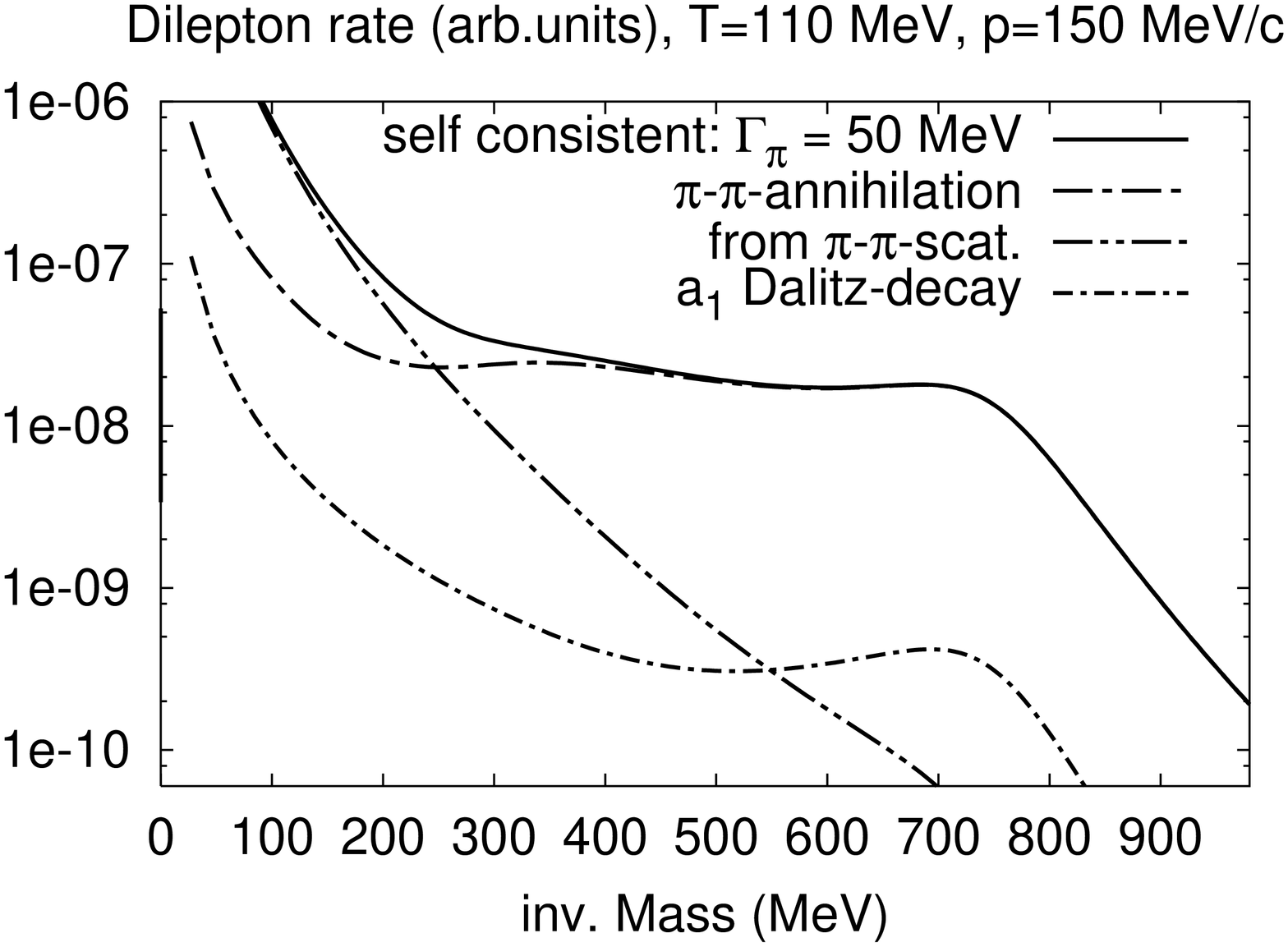}}}

\end{picture}
\caption{\textit{top: $\rho$-meson spectral function, bottom: thermal
    di-lepton rate.}}
\end{figure}

Self consistent equilibrium calculations are performed keeping the
full dependence of all two-point functions on three momentum ${\vec
  p}$ and energy $p_0$, and treating all propagators with their
dynamically determined widths. For simplicity the real parts of the
self energies were dropped and all time components of the polarization
tensors $\Pi_{\rho}$ and $\Pi_{a_1}$ were put to zero for reasons
discussed in sect. 3. The examples shown refer to a temperature of
$T=110 \,\text{MeV}$ appropriate for the CERES data. We discuss three
different settings. In case (a) the $\rho$-meson polarization tensor
is calculated simply by the perturbative pion loop, i.e. with vacuum
pion propagators and thermal Bose-Einstein weights (no self consistent
treatment). The two other cases refer to self consistent solutions of
the coupled Dyson scheme, where the four-$\pi$ interaction is tuned
such that the sun-set diagram provides a moderate pion damping width
of about $50\,\text{MeV}$ (case (b)) and a strong one of
$125\,\text{MeV}$ (case (c)) around the peak in the spectral function,
c.f.  fig.~1. Since in the thermal case any excitation energy is
available, though with corresponding thermal weights, all thresholds
disappear and the spectral functions show strength at \emph{all}
energies\footnote{In mathematical terms: all cuts go from $-\infty$ to
  $+\infty$ in energy, and the physical sheet of the retarded
  functions are completely separated from the physical sheet of the
  corresponding advanced function by these cuts.}!  The pion functions
shown in Fig. 1 are plotted against energy in order to illustrate that
there is significant strength in the space-like region (below the
light cone at $150\,\text{MeV}$) resulting from $\pi$-$\pi$ scattering
processes.

As an illustration we display a 3-d plot of the rho-meson spectral
function as a function of $p_0$ and ${|\vec p\,|}$ in Fig. 2, top
left.  The right part shows the spectral function as a function of
invariant mass at fixed three momentum of $150\,\text{MeV}/c$ in
vacuum and for the self consistent cases (a) to (c). The minor changes
at the low mass side of the $\rho$-meson spectral function become
significant in the di-lepton yields given in the left bottom panel.
The reason lies in the statistical weights together with additional
kinematical factors $\propto m^{-3}$ from the di-lepton decay
mechanism. For the moderate damping case
($\Gamma_{\pi}=50\,\text{MeV}$) we have decomposed the di-lepton rate
into partial contributions associated with $\pi$-$\pi$ bremsstrahlung,
$\pi$-$\pi$ annihilation and the contribution from the $a_1$-meson,
which can be interpreted as the $a_1$ Dalitz decay.

The low mass part is completely dominated by pion bremsstrahlung
contributions (like-charge states in the pion loop). This
contribution, which vanishes in perturbation theory is {\em finite}
for pions with finite width.  It has to be interpreted as
bremsstrahlung, since the finite width results from collisions with
other particles present in the heat bath. Compared to the standard
treatment, where the bremsstrahlung is calculated independently of the
$\pi$-$\pi$ annihilation process, this self-consistent treatment has a
few advantages. The bremsstrahlung is calculated consistently with the
annihilation process, it appropriately accounts for the
Landau-Pomeranchuk suppression at low invariant masses \cite{knoll96}
and at the same time includes the in-medium pion electromagnetic
form-factor for the bremsstrahlung part. As a result the finite pion
width adds significant strength to the mass region below
$500\,\text{MeV}$ compared to the trivial treatment with the vacuum
spectral function. Therefore the resulting di-lepton spectrum
essentially shows no dip any more in this low mass region already for
a moderate pion width of $50\text{MeV}$. The $a_1$ Dalitz decay
contribution given by the partial $\rho$-meson width due to the
$\pi$-$a_1$ loop in $\Pi_{\rho}$ is seen to be unimportant at all
energies. The present calculations have not included any medium
modification of the masses of the mesons. The latter can be included
through subtracted dispersion relations within such a consistent
scheme.

\section{Symmetries and gauge invariance}

While scalar particles and couplings can be treated self-consistently
with no principle problems at any truncation level, considerable
difficulties and undesired features arise in the case of vector
particles. The origin lies in the fact that, though in
$\Phi$-derivable Dyson re-summations symmetries and conservation laws
are fulfilled at the expectation value level, they are generally no
longer guaranteed at the correlator level. In the case of local gauge
symmetries the situation is even worse, because the symmetry of the
quantized theory is not the original one but the non-linear BRST
symmetry \cite{brs76,tyfr72}. Contrary to perturbation theory, where
the loop expansion corresponds to a strict power expansion in $\hbar$
and symmetries are maintained order by order, partial re-summations
mix different orders thus violating the corresponding symmetries.  It
is obvious that the scheme discussed above indeed violates the Ward
identities on the correlator level and thus the vector meson
propagators are no longer 4-dimensionally transverse.  This means that
unphysical states are propagated within the internal lines of the
$\Phi$-derivable approximation scheme which lead to a number of
difficulties in the numerical treatment of the problem. In the above
calculations we have worked around this problem by putting the
temporal components of the $\rho$-meson polarization tensor to zero,
an approximation, which is exact for ${\vec p}_{\rho}=0$.

Is there a self-consistent truncation scheme, where gauge invariance
is maintained also for the internal dynamics, i.e. for the dynamical
quantities like classical fields and propagators which enter the
self-consistent set of equations? The answer is definitely yes.
However one has to restrict the coupling of the gauge fields to the
expectation values of the vector currents. This in turn implies that
gauge fields are treated on the classical field level only, a level
that is presently explored in all hard thermal loop (HTL) approaches
\cite{Braaten91,Pisarski91,Blaizot93,Jackiw93}.  In the case of a
$\pi$-$\rho$-meson system the corresponding $\Phi$-derivable scheme is
then given by (again omitting the tadpole term)
\begin{eqnarray}
\label{piextrho}
\Phi\{G_{\pi},\rho\}&=&\hspace*{2mm}\GPhijmu\;\,+\hspace*{5mm}\GPhipi \\
\label{Sig-pi-cl}
\Sigma_{\pi}&=&\hspace*{7mm}\GSigjmu\hspace*{7mm}+\;\GSigpi\\
\label{cl-field-eq}
\left(\partial^{\nu}\partial_{\nu}-m^2\right) \rho^{\mu}&=&
\hspace*{-2mm}j^{\mu}=\;\Gjmu
\end{eqnarray}
Here full lines represent the self-consistent pion propagators and
curly lines with a cross depict the classical $\rho$-meson field,
governed by the classical field equations (\ref{cl-field-eq}). Since
$\Phi$ is invariant with respect to gauge transformations of the
classical vector field $\rho^{\mu}$, the resulting equations of motion
are gauge covariant. This also holds for fluctuations
$\rho^{\mu}+\delta \rho^{\mu}$ around mean field solutions of
(\ref{Sig-pi-cl} - \ref{cl-field-eq}). In this background field method
one can define a gauge covariant \emph{external} polarization tensor
via variations with respect to the background field $\delta
\rho^{\mu}$
\begin{equation}\label{Pi-ladder}
\Pi^{\text{ext}}_{\mu \nu}(x_1,x_2)=
\frac{\delta}{\delta \rho^{\mu}(x_2)}
\left [ \frac{\delta \Phi[G_{\pi},\rho]}{\delta  \rho^{\nu}(x_1)}
\right ]_{G_{\pi}=G_{\pi}[\rho]} =\; \PiVert
\end{equation}
as a response to fluctuations around the mean field. In order to
access this tensor one has to solve a corresponding three-point-vertex
equation
\begin{equation}
\label{vert-ladder}
\funcd{G_{\pi}}{\rho^{\mu}}=\;\GVert\;=\;\GVertV\;+\;\;\GVertLp
\end{equation}
In order to maintain all symmetries and invariances the Bethe-Salpeter
Kernel in this equation has to be chosen consistently with the
$\Phi$-functional (\ref{piextrho}), i.e.
\begin{equation}
\label{ladder-kernel}
K_{1234}=\frac{\delta^2\Phi}{\delta G_{12}\delta G_{34}}
=\;\GKVert
\end{equation}
Thereby the pion propagator entering the ladder resummation
(\ref{ladder-kernel}) is determined by the self-consistent solution of
the coupled Dyson and classical field equations (\ref{Sig-pi-cl} -
\ref{cl-field-eq}). Thereby the ladder re-summation also accounts for
real physical scattering processes. This phenomenon was already
discussed in \cite{knoll96} for the description of Bremsstrahlung
within a classical transport scheme (Landau-Pomeranchuk-Migdal
effect). From this point of view one clearly sees that the pure
$\Phi$-functional formalism without the vertex corrections provided by
(\ref{vert-ladder}) describes only the ``decay of states'' due to
collision broadening. Thus the \emph{internal} polarization tensor
given in the $\Phi$-Dyson scheme (\ref{Phi_Dyson}) has a
\emph{time-decaying} behavior, with the 00-component approximately
behaving like
\begin{equation}
\label{Pi00-time1}
\Pi^{00}_{\rho}(\tau,{\vec p}=0)\propto e^{-\Gamma \tau}
\end{equation}
in a mixed time-momentum representation. This clearly violates charge
conservation, since $\partial_0\Pi^{00}_{\rho}(\tau,{\vec p})|_{{\vec
    p}=0}$ does not vanish! Accounting coherently for the multiple
scattering of the particles through the vertex re-summation
(\ref{vert-ladder}) keeps track of the ``charge flow'' into other
states and thus restores charge conservation.  Within classical
considerations the ladder re-summation (\ref{Pi-ladder}) indeed yields
\begin{equation}
\label{Pi00-time2}
\Pi^{00}_{\rho}(\tau,{\vec p}=0)\propto 
\sum_n \frac{(\Gamma\tau)^n}{n!}\; e^{-\Gamma \tau}=1
\end{equation}
confirming charge conservation. For further details c.f. ref. \cite{knoll96}.

Unless one solves the exact theory there seems to be no obvious self
consistent alternative to the above scheme where vector particles are
treated dynamically and which at the same time complies with gauge
invariance also for the internal propagation. All this seems to defer
a dynamical treatment of vector particles on the propagator level. The
numerical implementation of the above vertex corrections is in
progress. The problem of renormalization omitted here has been
investigated using subtracted dispersion relations.  Thus for vector
particles a fully self-consistent scheme with all the features of the
$\Phi$-functional, especially to ensure the consistency of dynamical
and thermodynamical properties of the calculated propagators together
with the conservation laws on both the expectation value and the
correlator level remains an open problem.

The equilibrium calculations presented also serve the goal to gain
experience about particles with broad damping width with the aim
towards a transport scheme for particles beyond the quasi-particle
limit \cite{ikv99-2},
see also \cite{Leupold99,Cassing99}.\\[-10mm]

\end{fmffile}

\end{document}